\documentclass[useAMS,usenatbib]{mn2e}
\usepackage{graphicx}

\title{Non-circular motion evidences in the circumnuclear region of M100 (NGC 4321)}

\author[A. Castillo-Morales, J. Jim\'enez-Vicente, E. Mediavilla, E.
Battaner]{A. Castillo-Morales$^{1}$\thanks{E-mail:acm@astrax.fis.ucm.es},
J. Jim\'enez-Vicente$^{2}$, E. Mediavilla$^{3}$, E. Battaner$^{2}$
\\
$^{1}$Dpto. de Astrof\'{\i}sica y CC. de la Atm\'osfera, Universidad
Complutense de Madrid, Spain.\\
$^{2}$Dpto. F\'{\i}sica Te\'orica y del Cosmos, Universidad de
Granada, Spain.\\
$^{3}$Instituto de Astrof\'{\i}sica de Canarias, Tenerife, Spain.\\}

\begin{document}

\date{}


\maketitle


\begin{abstract}
We analyse new integral field spectroscopy of the inner region (central
2.5 kpc) of the spiral galaxy NGC 4321 to study the peculiar kinematics
of this region.
Fourier analysis of the velocity residuals obtained by subtracting an
axisymmetric rotation model from the $\rm H\alpha$ velocity field, indicates
that the distortions are {\em global} features generated by an $m=2$
perturbation of the gravitational potential which can be explained by
the nuclear bar.
This bar has been previously observed in the near-infrared but not in
the optical continuum dominated by star formation. We detect the optical
counterpart of
this bar in the 2D distribution of the old stellar population (inferred
from the equivalent width map of the stellar absorption lines).
We apply the Tremaine--Weinberg method to the stellar velocity field to
calculate the
pattern speed of the inner bar, obtaining a value of
$\Omega_b$=160$\pm70\rm \ km \ s^{-1}\ kpc^{-1} $. This value is considerably larger than the one obtained when a simple bar model is considered.
However the uncertainties in the pattern speed determination prevent us to give support to alternative scenarios.

\end{abstract}


\section{Introduction}

Departures from regular rotation, such as winds, elongated orbits or
streaming motions, provide very relevant information about
the dynamics and history of galaxies. To identify kinematical
distortions and to infer how they relate to the gravitational
potential, different techniques such as velocity field fitting or
Fourier analysis can be used. However, to obtain reliable results
these tools must be used in combination with 2D kinematical data of good
spatial resolution. Integral field spectroscopy (IFS) is
an observational technique especially suited for kinematical studies that
can be applied to nearby galaxies, such as M100 (NGC 4321),
to obtain velocity maps with detailed spatial information.

The inner region of NGC 4321 has been extensively studied at different
wavelengths. Optical~\citep{Pierce86}, H$\alpha$,
near-infrared~\citep{knapen95a,knapen95b} and CO
studies~\citep{Rand95,Saka95,sempere97},
have revealed some interesting irregular features.
On the one hand, the inner region of this galaxy shows a peculiar
morphology that is quite
different in the optical and the
NIR. In the optical, the structure is dominated by two spiral arms
(broad band)
and an ovally shaped region of enhanced
star formation ($ \rm H\alpha,\ H\beta$), while the near-infrared shows
an inner
bar aligned with the large scale
stellar bar and a pair of small arms emerging from its
ends~\citep{knapen95a,knapen95b}.
Whether there is a single bar or two nested bars that just happen to be
aligned
has been a greatly debated
issue~\citep{knapen95b,knapen00,wada98,Burillo98}. On the other hand,
~\citet{knapen00} found non-circular motions from H$\alpha$ data, and
interpreted them as the
kinematical signatures of gas streaming along the inner part of the bar,
and of
density-wave streaming motions across a
two-armed minispiral. Relevant pattern speeds of this galaxy have been measured at larger scale by different methods~\citep[and references therein]{canzian97,sempere95,hernandez05}.
Recently, ~\citet{allard05a} have analysed the
inner region of NGC 4321 from IFS observations obtained with
SAURON. They focus on the star formation ring, which they confirm
to be formed by young stars, but find a low gas velocity dispersion
compared to
its surroundings.

In this paper we present new IFS (covering a large
wavelength range in the optical) obtained with INTEGRAL, a system of
exchanging
fibre bundles at the WHT suited for this
kind of study~\citep{mediavilla05,battaner03,glorenzo01}.
We  use the stellar and ionized gas velocity fields to analyse the
symmetries of the departures from pure rotation and to estimate the
nuclear bar pattern speed by means of the Tremaine \& Weinberg (1984)
method. These studies will be used to relate the
morphological and kinematical distortions observed in the inner 2.5 kpc
with the
gravitational potential of the nuclear bar.

NGC 4321 has a very accurately (Cepheid) measured distance of 16.1
Mpc~\citep{Ferrarese96} which
corresponds to 78 pc arcsec$^{-1}$ on the plane on the sky. We use this
value
throughout this work.

The main properties of M100 are summarized in Table~\ref{latabla}.

\begin{table}
\label{latabla}
  \begin{tabular}[h]{ll}
    \hline
    Name &  M100 \\
    Type & SAB(s)bc LINER-HII \\  
    R.A.  (J2000) & $12^h 22^m 54.9^s$\\
    Dec. (J2000) & $15^\circ 49' 20''$ \\
    Incl. &  $30^\circ$ \\
    P.A. &  $151^\circ \pm 3^\circ$ \\
    Total Mag.  (B) & 10.05 \\
    Sys. Velocity (opt) &  1567 $\pm$ 7 km/s \\
    Distance & 16.1 Mpc \\
    Linear scale & 78 pc/arcsec \\
    \hline
  \end{tabular}
\caption{M100 main properties}
\end{table}

\section{Observations and data reduction}
\label{section:obs}

\begin{figure*}
   \centering
   \includegraphics[angle=0,width=8.cm, clip=true]{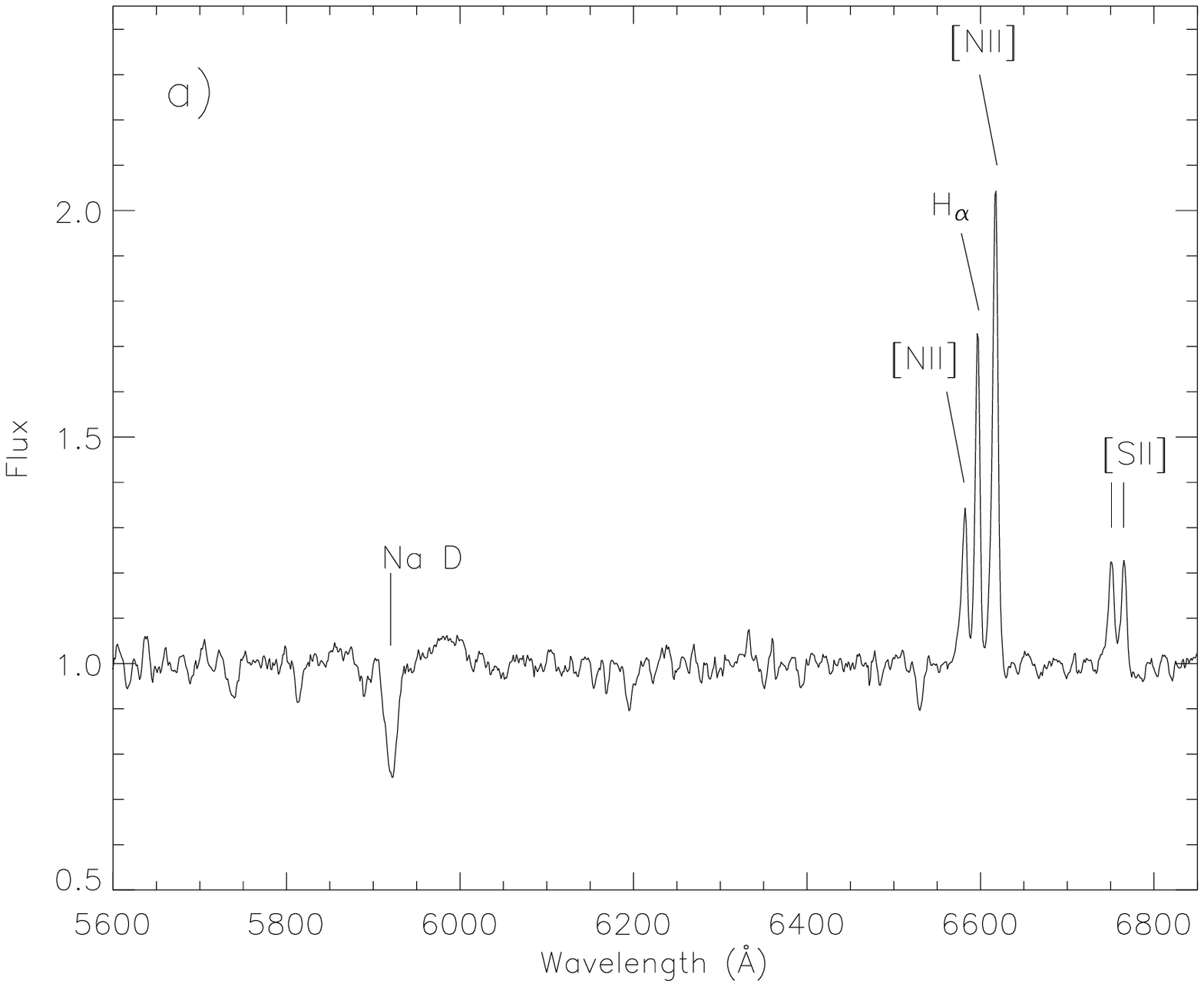}
   \includegraphics[angle=0,width=9.5cm, clip=true]{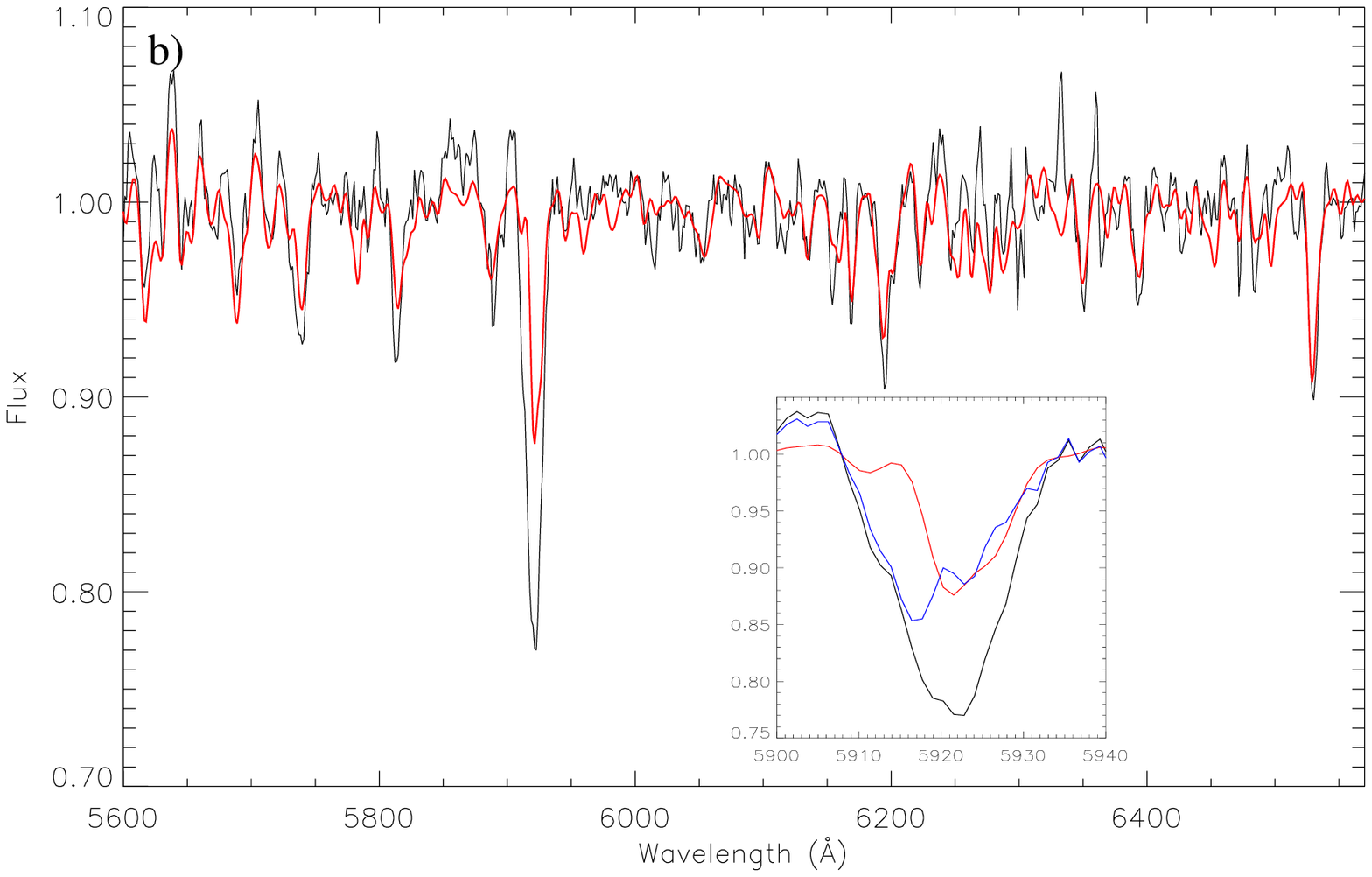}
   \caption{{\bf{(a)}} Nuclear spectrum (fibre $\#$68 in bundle SB3) of
NGC 4321
in the range 5600--6850 \AA\ showing the most important absorption and
emission
lines. {\bf{(b)}} Nuclear spectrum for the SB3 fibre bundle with the
stellar
template plotted on top in red line. This template is used with the
XCSAO
task to derive the stellar velocity field using the cross-correlation
technique
in the wavelength range 5600--6560\AA. The inner panel shows in detail
the fit of
the Na D absorption doublet (red line) and the residuals (blue line).}\label{figure:sp_gx}
\end{figure*}

The data analysed in this article were obtained on 2002 March 16 at the
Observatorio del Roque de los Muchachos on the island of La Palma with
the fibre
system INTEGRAL~\citep{Arribas98} in combination with the fibre
spectrograph
WYFFOS at the William Herschel Telescope~\citep{Bingham94}. The weather
conditions during this night were fairly good, with a seeing of about
1.3$\arcsec$. The data discussed in this paper were obtained with INTEGRAL
standard bundles $\#$3 and $\#$2. The WYFFOS spectrograph was equipped
with a
1200 groove $\rm mm^{-1}$ grating centred on 6247 \AA. The spectral
resolution
was 4.8\AA\ (R $\approx$ 1300) for SB3 and 2.8\AA\ (R $\approx$ 2200) 
for SB2 fibre bundles. INTEGRAL+WYFFOS
allow us to observe a large (compared with other
IFS systems) spectral range (5600-6850\AA) that, in addition to
H$\alpha$, also contains the high excitation [N
II]$\lambda\lambda$6548, 6584 and [S II]$\lambda\lambda$6716, 6731,
emission lines and the Na D $\lambda\lambda$5890, 5896 absorption
features (see
Fig.~\ref{figure:sp_gx} (a)) not included in previous IFS studies.
With this configuration, and pointing to the centre of NGC 4321, we took
three
exposures of 1200 s each during the night.

The data reduction of fibre-based 2D spectroscopy~\citep[
e.g.][]{Arribas91,Mediavilla92} consists of two main steps: i) basic
reduction
of the spectra (i.e.\ bias, flat-fielding, extraction, wavelength
calibration,
etc.) and ii) generation of maps of spectral features (e.g.\ line
intensity,
velocity fields, etc.) from the spectra. Step (i) was performed in the IRAF
environment following standard procedures. We obtained typical wavelength
calibration errors of 0.15 \AA, which give velocity uncertainties of
$\pm$ 7
$\rm km \ s^{-1}$ for H$\alpha$. For step (ii) we have developed our own
software
packages. In particular we transform an ASCII file with the actual
position of
the fibres and the spectral feature corresponding to each fibre into a
regularly spaced rectangular grid. In this way we build up images of 33
$\times$
29 pixels for SB3 bundle, and 45 $\times$ 34 pixels for SB2, with a
scale of
$\sim 0.95\arcsec \rm pixel^{-1}$, and $0.35\arcsec \rm  pixel^{-1}$ for
SB3 and
SB2 fibre bundles respectively. The resulting images are therefore of 
sizes $31.35''\times 27.55''$ for SB3 and $15.75'' \times 11.9''$ for SB2.
These images can be treated with standard
astronomical software.

\section{Results}
\label{section:results}

\subsection{Nuclear Spectrum}

The stellar population in the central region of NGC 4321 has been carefully
analysed by~\citet{Sarzi05}. They estimate an average age of
$~$1 Gyr at solar or super-solar metallicities, with only a
marginal improvement in the results for super-solar metallicities.
We therefore use a synthetic stellar population of 1 Gyr with solar
metallicity
as a template for the stellar absorption feature analysis. We have used the
synthetic population by~\citet{Delgado05}, which covers the observed
wavelength
range at a very high spectral resolution (0.3 \AA).
This range includes the strong NaD line that might be affected by
interstellar
absorption. We thus exclude the wavelength range of NaD from the fitting,
obtaining a template that reproduces very well the observed absorption
features,
except the NaD line, which exhibits strong interstellar contamination. The
difference between the nuclear spectrum and the template shows that the
interstellar contribution is blue-shifted (see Fig.~\ref{figure:sp_gx}
(b)).
This is a  notable result, since blue-shifted velocity components in this
feature
unambiguously indicate the presence of outflowing gas, as demonstrated in
previous studies~\citep[see][and references therein]{rupke05}.

\subsection{H$\alpha$ Maps}

\begin{figure*}
   \centering
\includegraphics[width=\textwidth, clip=true]{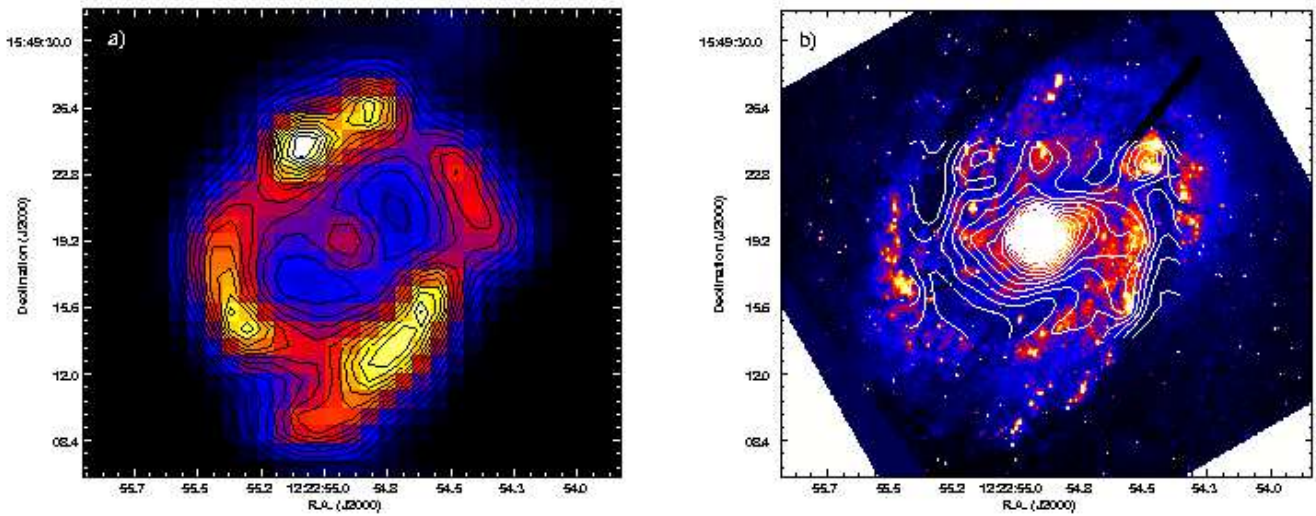}
\caption{{\bf{(a)}} H$\alpha$ intensity map computed for the SB3 fibre
bundle.
The H$\alpha$ emission is distributed in a nucleus and an ovally shaped
ring.
{\bf{(b)}} HST/ACS (F555W) image of NGC 4321 with white contours
overlaid from
the continuum map computed for the SB2 fibre bundle.}
   \label{figure:intha_cont}
\end{figure*}

To obtain the intensity and velocity maps for
the ionized gas we fit a single gaussian to the
H$\alpha$ emission line after subtracting the stellar contribution from each
spectrum.  Absorption correction is in general not very
important except
at the nucleus. According to the H$\alpha$ intensity map (see
Fig.~\ref{figure:intha_cont} (a)), the emission is distributed in a
nucleus and
an ovally shaped ring~\citep{knapen00,allard05a}, where several HII
regions can
be distinguished.

\begin{figure*}
\centering
\includegraphics[angle=0,width=17cm, clip=true]{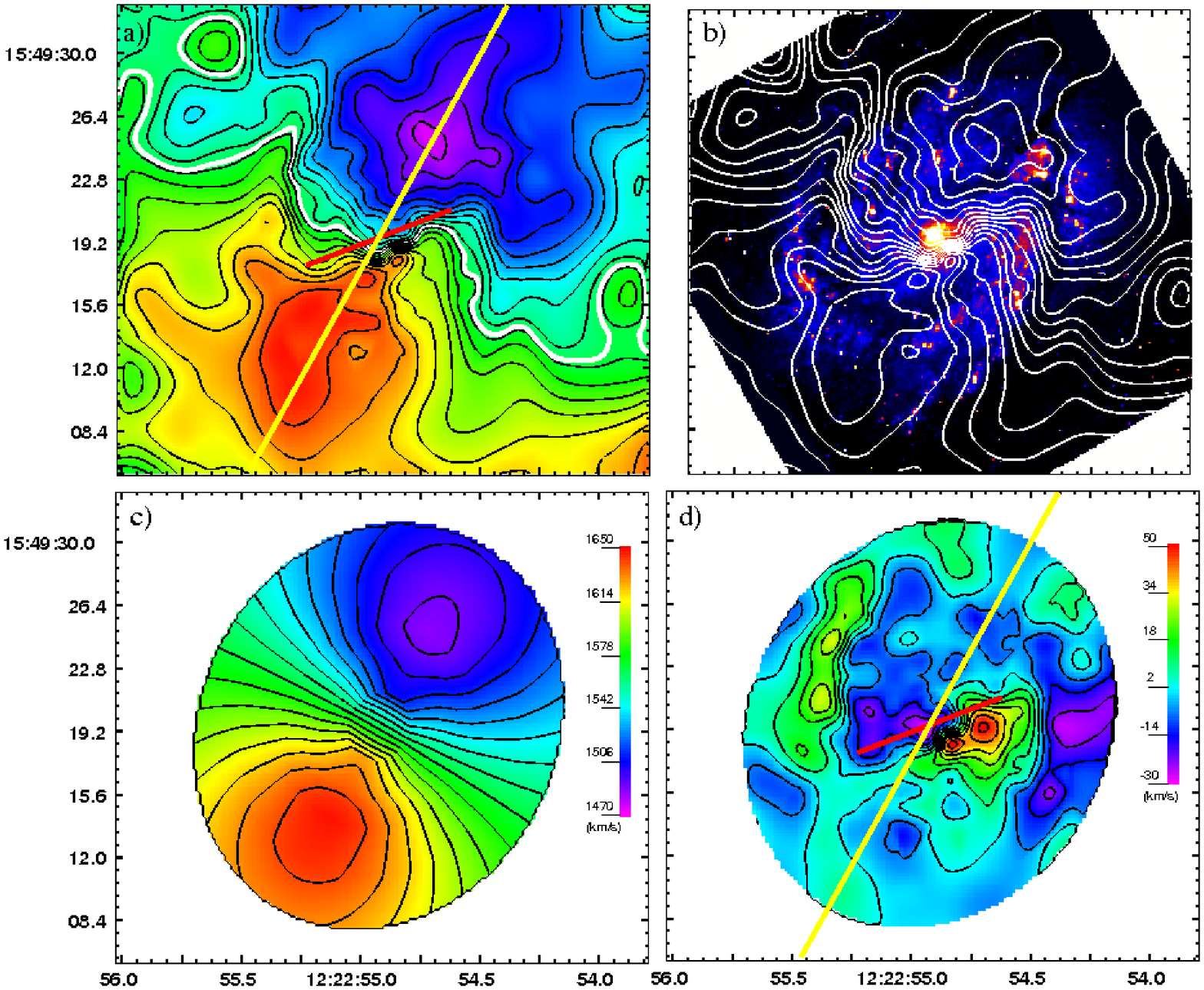}
\put(-360.,-8.){R.A~(J2000)}
\put(-140.,-8.){R.A~(J2000)} 
\put(-470.,65.){\rotatebox{90}{Declination~(J2000)}}
\put(-470.,260.){\rotatebox{90}{Declination~(J2000)}}
\vspace{0.2cm}
\caption{{\bf{(a)}} H$\alpha$ velocity map computed with the information
from
the SB2 and SB3 fibre bundles. Contour levels range from 1470 $\rm km \
s^{-1}$ to 1650 $\rm km \ s^{-1}$ in steps of 10 $\rm km \ s^{-1}$.
The thick white contour indicates the systemic velocity at 1567 $\rm km \
s^{-1}$. Figures (a) and (c) share the same scale colour bar.
{\bf{(b)}} HST/ACS (F555W) image of NGC 4321 with white contours of
H$\alpha$
velocity field overlaid. {\bf{(c)}} Model velocity map with fixed PA
(see text).
{\bf{(d)}} Map of residual velocities derived by subtracting the model
velocity
field (c) from the observed velocity map (a). Contour levels: from $-30$
$\rm km \
s^{-1}$ to 50 $\rm km \ s^{-1}$ in steps of 8 $\rm km \
s^{-1}$.
Yellow and red lines show the position angle of the galaxy (P.A =
$151^\circ$)
and the inner stellar bar (P.A = $111^\circ$) respectively.}
\label{figure:kin}
\end{figure*}

The H$\alpha$ velocity field map is computed with the combined
information from
the SB2 and SB3 fibre bundles (see Fig.\ref{figure:kin} (a)). Our
velocity map looks quite similar to the one published
by~\citet{knapen00}. It
essentially shows a basic rotational pattern
significantly distorted by the effect of the inner bar and spiral arm
structures. In Fig.\ref{figure:kin} (b) we overplot the ionized gas
velocity
contours on the HST/ACS (F555W) image. In this
figure, the two regions with clear departures from axisymmetric circular
rotation
stand out clearly. The innermost region shows a clear twist in the
kinematic
axes due to the influence of the inner bar. Further out, the
deviations follow the spiral arms.

\subsection{Stellar Maps}

\begin{figure*}
\centering
\includegraphics[angle=0,width=8cm, clip=true]{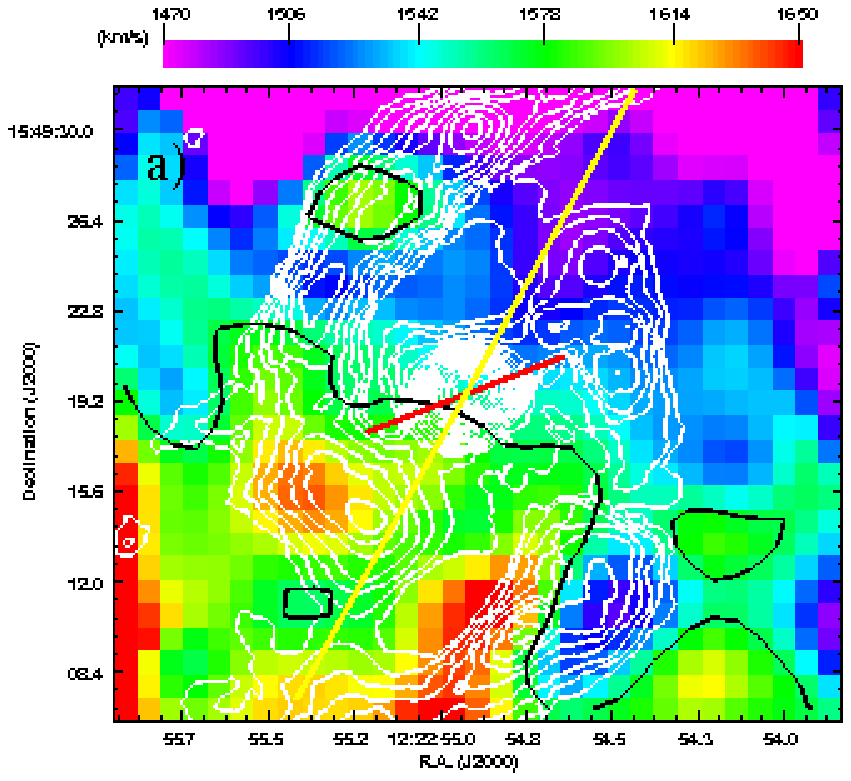}
\includegraphics[angle=0,width=8cm, clip=true]{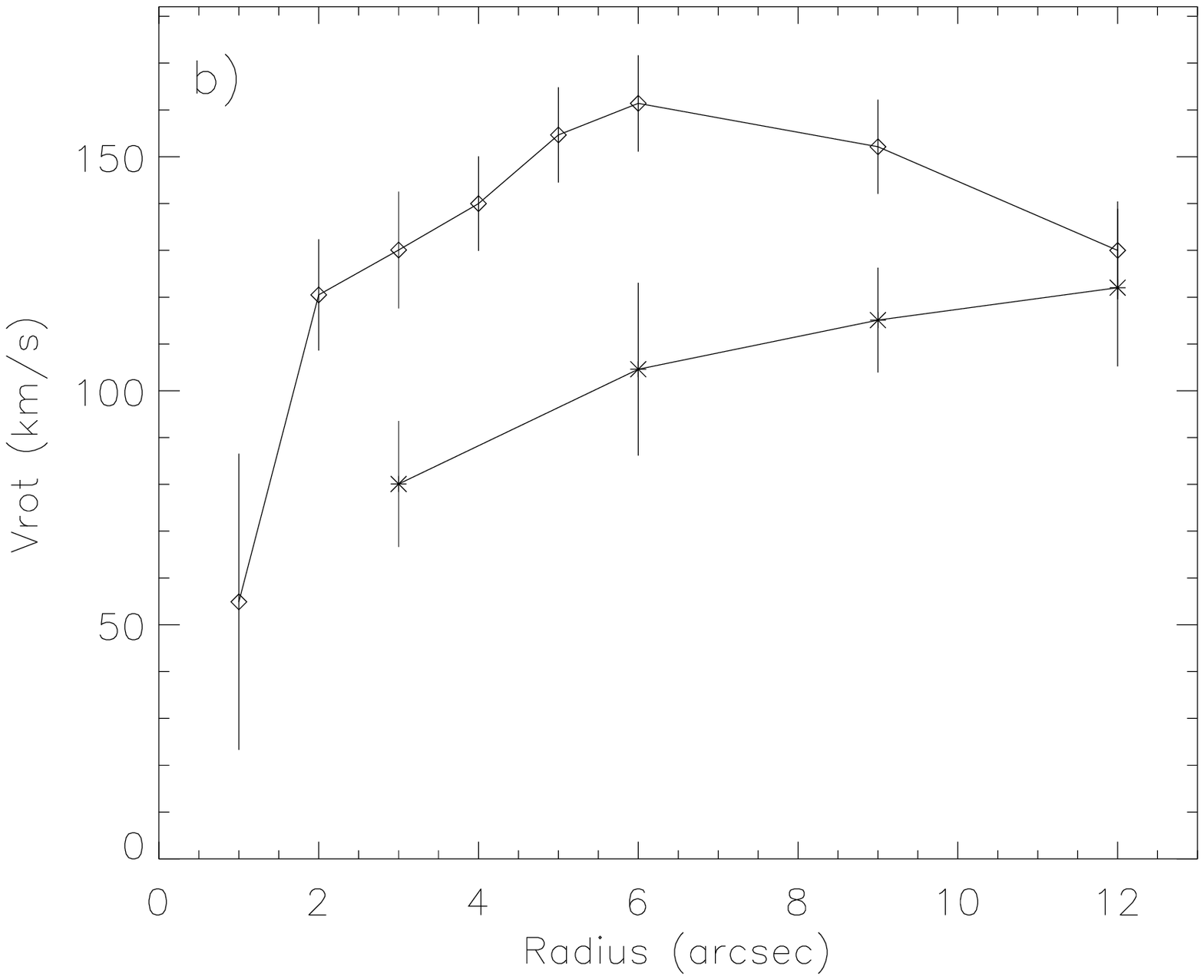}
\caption{{\bf{(a)}} Stellar velocity field for the SB3 fiber bundle.
Velocities range from 1470~$\rm km \ s^{-1}$ to 1650 $\rm km \
s^{-1}$.
The thick black contour indicates the systemic velocity at 1567 $\rm km \
s^{-1}$. Yellow and red lines show the position angle of the galaxy (P.A. =
$151^\circ$) and the inner stellar bar (P.A. = $111^\circ$) respectively.
Contours of the CO emission are overplotted in white.
{\bf{(b)}} Stellar (star symbols) and $\rm H\alpha$ (diamond symbols)
rotation
curves of the inner disk of NGC 4321 derived up to $12\arcsec$ ($\approx
1$ kpc)
from the galactic centre.}
\label{figure:vel_stellar}
\end{figure*}

Cross-correlation of the spectra is performed in the
5600--6560\AA~wavelength
range with the XCSAO task from the IRAF package.
We have excluded in the cross-correlation wavelength ranges around the [O
I]$\lambda\lambda$6300, 6363 lines to avoid residual sky emission
contamination
and
a wavelength range around the NaD feature (for the reasons mentioned
above). The
stellar velocity uncertainties ranges from $\rm 15 \ km \ s^{-1}$ (inner
regions) to $\rm 35 \ km \ s^{-1}$ (outer regions).  The resulting stellar
velocity field is shown in Fig.~\ref{figure:vel_stellar} (a).
It looks very similar to the results published by~\cite{allard05b}.
Although noisier than the gas velocity field, it shows a stellar rotation
pattern that is more regular than the velocity pattern of the ionized
gas. It shows some distortions in
the bar region although not as strong as in the case of the gas. The
strongest
deviations
from circular rotation in the map take place at two spots in the spiral
arms.
They are located roughly symmetrically with respect to the galactic
centre with
a P.A. of about $45^\circ$ from the major axis (nearly perpendicular to
the bar
P.A.), coincident with the highest density regions of the CO map
by~\citet{Saka95}. This supports the role of the nuclear bar in explaining
the complex kinematics and morphology in the innermost region of NGC
4321. This
issue, and the question of whether the nuclear bar is part of the large
scale
one present in this galaxy or a separated structure~\citep[see][for a
review]{hernandez05} will be addressed in Section~\ref{section:stelkin}.

The stellar continuum map shows very similar features to the broad band
HST/ACS
(F555W) image although
 at a lower spatial resolution (see Fig.~\ref{figure:intha_cont} (b)).
In this
map we can distinguish
the nucleus emission from which a spiral structure is coming out in the SW
direction and a fainter one at the NE direction. Dust structures at the
inner
rims of the arms are also clearly seen in the INTEGRAL continuum
map~\citep[see
also][]{knapen95a, knapen95b}.
The brightest regions in this map trace stellar formation regions, of
which is
particularly clear the bright region surrounded by dust in the NW of the
nucleus.

\begin{figure}
\centering
\includegraphics[width=8.cm, clip=true]{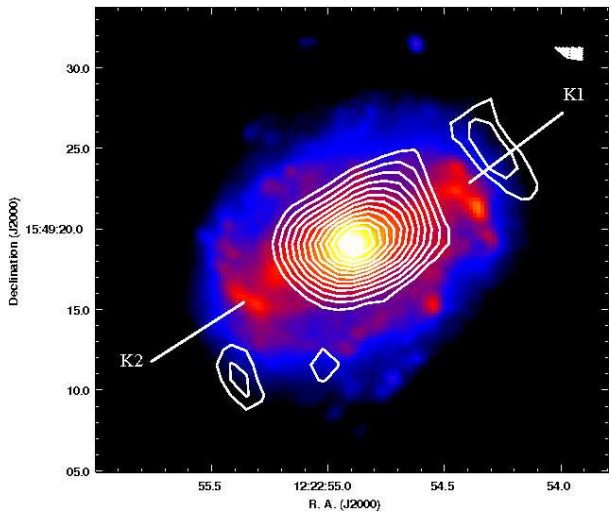}
\caption{Infrared ($K_{\rm s}$) image of NGC 4321 from~\citep{knapen03}
with contours of the NaD equivalent width for the SB3 fibre bundle.
Contour levels range from 0.75 to 1.3\AA~in steps of 0.05\AA. {\bf{Infrared knots K1 and K2~\citep{knapen95a} are marked.}}}
\label{figure:ewsb3na}
\end{figure}

We have also calculated maps for the strength and equivalent width of the
stellar absorption lines from the fittings of the template to the spectra
obtained during the cross-correlation procedure (see previous section). It
should be kept in mind that we have used the
same single age, single metallicity population template for the whole
map, and that
therefore
the maps for the intensities and equivalent widths of all the stellar
absorption
lines are proportional. This is obviously an oversimplification, as we know
there will be variations in stellar populations throughout the field of
view,
but
a thorough analysis of the stellar populations is beyond the scope of this
paper, and these maps are intended only to be a tracer of the global
stellar
mass content of the galaxy. Notice that we have obtained a map normalized
(according to the template) to the NaD line absorptions, which are the
strongest
ones in the  wavelength range studied. This NaD map is therefore free
from interstellar contamination.

Figure~\ref{figure:ewsb3na} shows the contours of the equivalent width
of the
stellar NaD line on top of the WHT $K_{\rm }s$ image
from~\citet{knapen03}. The iso-levels of the
EW in the innermost region are elongated with a P.A. very close to the
111$^\circ$
quoted by~\citet{knapen95a} for the inner bar. The extent of the bar
does also
roughly agree with the 9 arcsec proposed by~\citet{knapen95a}. Although
this is
not completely
surprising (as the equivalent width is expected to be somewhat less
affected by
dust than continuum or broadband maps), it clearly shows that the EW is
a good
tracer of the old population responsible for the absorption features.
The two
well known infrared knots K1 and K2~\citep{knapen95a} do not show up in
the EW
map and in fact correspond to minima on it {\bf{(see Figure~\ref{figure:ewsb3na})}}. These knots are known to be
regions
of powerful recent star formation~\citep{ryder99, wozniak98} and are
therefore very
bright because of the contribution of young stars, but not necessarily
rich in
old stars responsible for the absorption features contributing to the
EW. Although the spectra are smeared by poor angular resolution, we have
been able to detect,
for the first time, the (otherwise hidden) inner bar through the study of
absorption spectral features characteristic of old stars. We would like to
point out that, the $K_{\rm s}$ band image shows very little hint of the
spiral
arms~\citep{knapen95a,knapen95b}. This fact indicates that the mass
distribution
is quite axisymmetric at radii larger than $\approx 6\arcsec$.

\section{Discussion}
\label{section:discussion}
\subsection{Residual Velocity Field of the Ionized Gas}
\label{section:gaskin}

Departures from circular motion are better addressed through residual
velocity
analysis. We  therefore start by deriving a pure rotation model by fitting
tilted rings to the observed velocity field~\citep{begeman89}. This is a
standard procedure for which the galaxy is divided into concentric
elliptical annuli where each annulus is characterized by the inclination
angle
($\rm i$), the position angle of the major axis (P.A.), the
rotational velocity ($\rm v_{c}$), the systemic velocity of the galaxy
($\rm
v_{sys}$) and the coordinates of the centre of each annulus
($\rm x_{c},y_{c}$). These parameters are fitted following a least squares
algorithm. We have used the task ROTCUR from the data reduction software
package
GIPSY (Groningen Image Processing SYstem) for this
purpose.

\begin{figure}
\centering
\includegraphics[angle=0,width=8cm, clip=true]{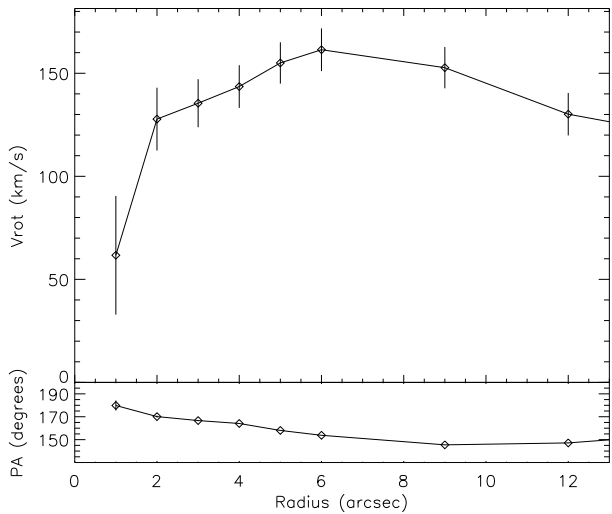}
\caption{H$\alpha$ rotation curve of the inner disc of NGC 4321 derived
up to
$12\arcsec$ ($\approx$ 1 kpc) from the galactic centre. The lower panel
shows
variations of P.A. with radius.}
\label{figure:rcgas}
\end{figure}

We take the kinematical centre at the position of the nuclear source at our
highest available
spatial resolution. This position matches, within the resolution limits,
the
position of the brightest spot in the HST image. We fix the inclination
to a
value of $30^{\circ}$~\citep{knapen02}. The systemic velocity is
calculated as
the average value of
the outermost rings. We obtain a value of 1567 $\pm$ 7 $\rm km \ s^{-1}$
. With
these values fixed, we perform a fit leaving the P.A. and the rotational
velocity of each ring as free parameters. The results are shown in
Figure~\ref{figure:rcgas}. The error bars in the rotation curves of 
Figures ~\ref{figure:rcgas} 
and ~\ref{figure:vel_stellar} b) are calculated from the difference in the 
rotation curves calculated for the approaching and receding sides of the 
galaxy and the error in the velocity determination, which have been added 
quadratically.

The  rotation curve obtained shows a very similar behaviour to previous
observations \citep{knapen00, Saka95}. Our values
are in fact somewhat lower than those reported by~\citet{knapen00}, but
in good
agreement with
CO observations by~\citet{Saka95}. Although, as usual, the inner region
provides
the most uncertain determination of the rotation curve, we have checked
that
rotations curves calculated for the maps at the
two available resolutions (SB2 and SB3 fibre bundles) give an excellent
agreement for the values of the rotational velocity in the overlapping
regions.
Resolution effects will therefore not be too relevant
in the shape of the inner rotation curve.

The rotation curve shows a very steep rise in the inner 3$\arcsec$,
reaching
values of 135 $\rm km \ s^{-1}$.
The velocity reaches a maximum of about 160 $\rm km \ s^{-1}$ at
6$\arcsec$ (470
pc) and then declines smoothly to a value of 130 $\rm km \ s^{-1}$ in the
outermost ring at 12$\arcsec$ ($\approx$ 1 kpc).

To investigate the non-circular motions in the galaxy disc we compute the
residual velocity map.
We would like to minimize the effect of non-circular motions being
absorbed in
the rotational model.
We use for this purpose a
rotational model with a fixed P.A. (= $151^\circ\pm 3^\circ$, which is
the average
of the
P.A. values corresponding to the outermost rings). In this way, most of the
information on
non-axisymmetric components is kept in the residual velocity. Of course,
deviations from circular rotation that take place along or near
the major axis in a symmetric way (on both the approaching and the
receding sides) will
most probably be absorbed in the rotation curve shape. There is little
hope of
disentangling the non-circular and circular components along the major
axis, and
therefore we adopt this as the best possible solution.
The calculated axisymmetric rotational model is displayed in
Figure~\ref{figure:kin} (c).
Subtracting this model from the observed velocity field, we obtain  
the residual velocity map for NGC 4321 which is shown in
Figure~\ref{figure:kin}
(d).

\begin{figure*}
\centering
\includegraphics[angle=0,width=8cm, clip=true]{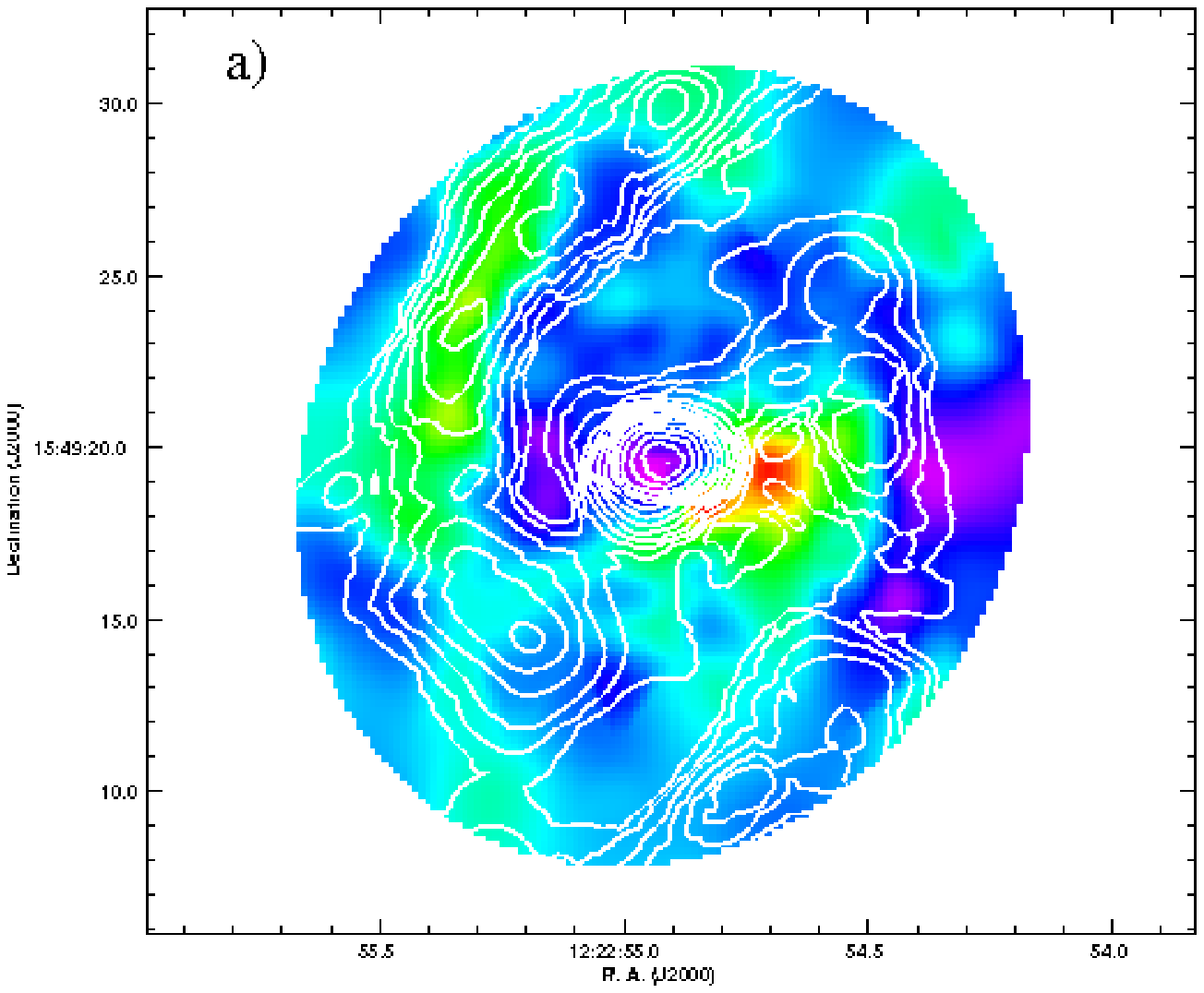}
\includegraphics[angle=0,width=8cm, clip=true]{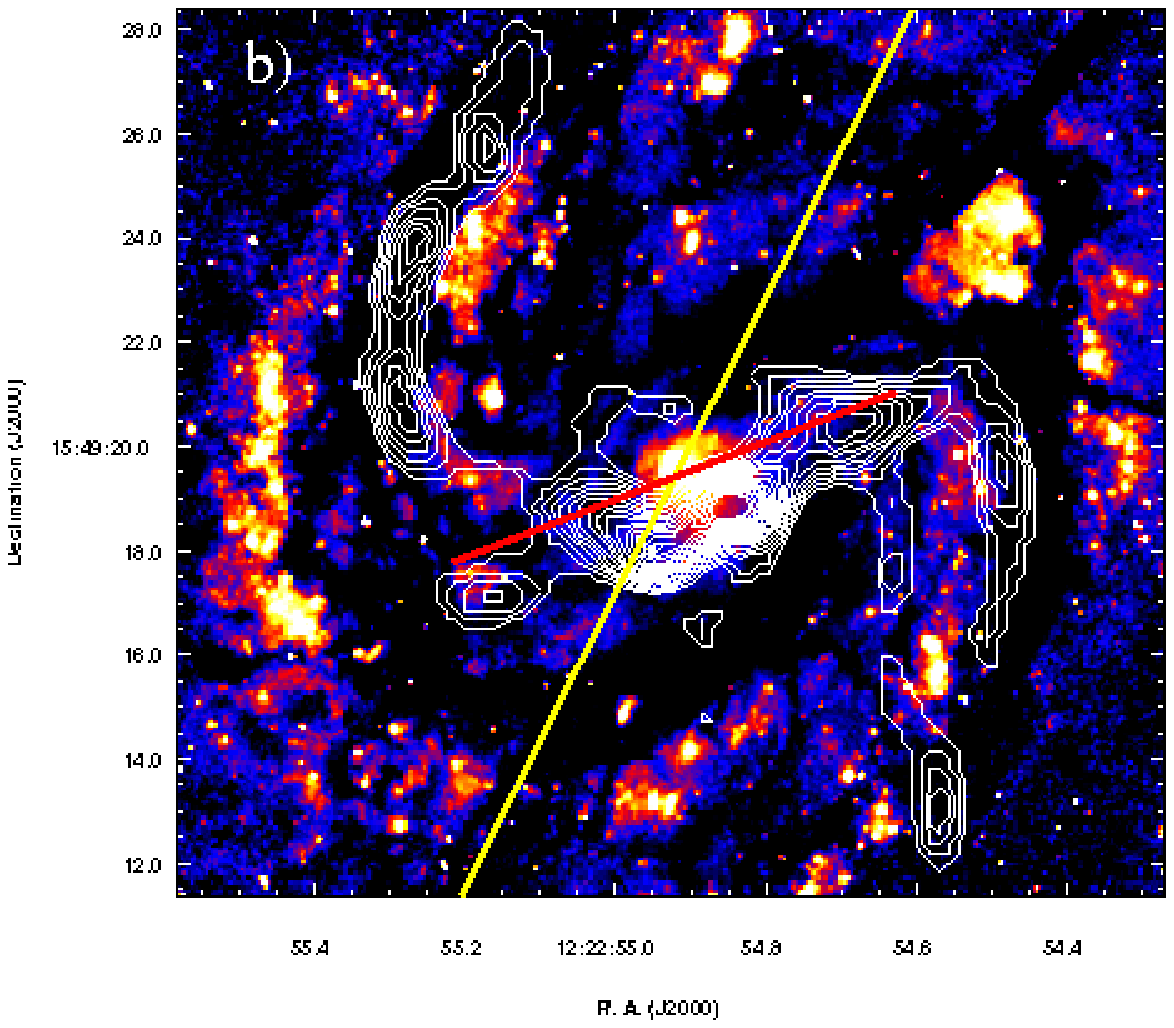}
\caption{{\bf{(a)}} The residual velocities (colour image) nicely match the
position of the molecular gas from the CO emission shown in white contours.
{\bf{(b)}} HST/ACS (F555W) unsharp masked image of NGC 4321 with residual
velocity gradient contours overlaid. Yellow and red lines show the position
angle of the galaxy (P.A. = $151^\circ$) and the inner stellar bar (P.A. =
$111^\circ$) respectively.}
\label{figure:resco_robdust}
\end{figure*}

Three regions can be clearly identified in the residual velocity map:
(i) A blue-shifted nuclear region that shows blue residuals of about
$-25$ $\rm km \
s^{-1}$. We would like to point out that although the rotation curve is
quite
uncertain in the innermost region, the existence of these blue residuals
in the
nuclear region does mostly depend on the position of the kinematic
centre and
on the systemic velocity, which are both pretty well constrained by the
observations. This fact suggests the existence of an outflow in this
region. (ii) A second region, further out up to radii of about
6$\arcsec$, that shows a
clear signature of influence of the nuclear bar (whose position angle
and size
are shown in Fig.~\ref{figure:kin} (d) by the red line). The map shows
residuals
of nearly 50 $\rm km \ s^{-1}$ on the west side
and of $-25$ $\rm km \ s^{-1}$ on the east side. These are the zones
where orbits
tend to align with the bar
potential. The symmetric shape of this region above and
below the bar is quite remarkable. (iii) At larger radii (r $ > 6
\arcsec$) the residual map shows clear signatures
of the presence of the spiral arms with values of 30 $\rm km \ s^{-1}$
for the
northern arm and $-25$ $\rm km \ s^{-1}$ for the southern arm.
The northern arm is more clearly depicted than the southern one. In
fact, as can be seen in Fig.~\ref{figure:resco_robdust} (a), the
residuals for the northern
arm  nicely match the position of the molecular gas from the CO map
by~\citet{Saka95}. The southern arm is somewhat displaced from the position
where the highest residuals are found. The residual velocities along the
arms
seem to vanish as they approach the major axis, although, as we have
mentioned
before, it is very likely that a large fraction of the non-circular
motions has been absorbed in the shape of the rotation curve.
The latter two of the three regions have already been identified
by~\citet{knapen00}. In Section~\ref{section:sym} we  perform a
Fourier analysis of the ionized gas kinematics to study the relationship
of these regions with the gravitational potential.

\begin{figure*}
\centering
\includegraphics[angle=0,width=7.9cm, clip=true]{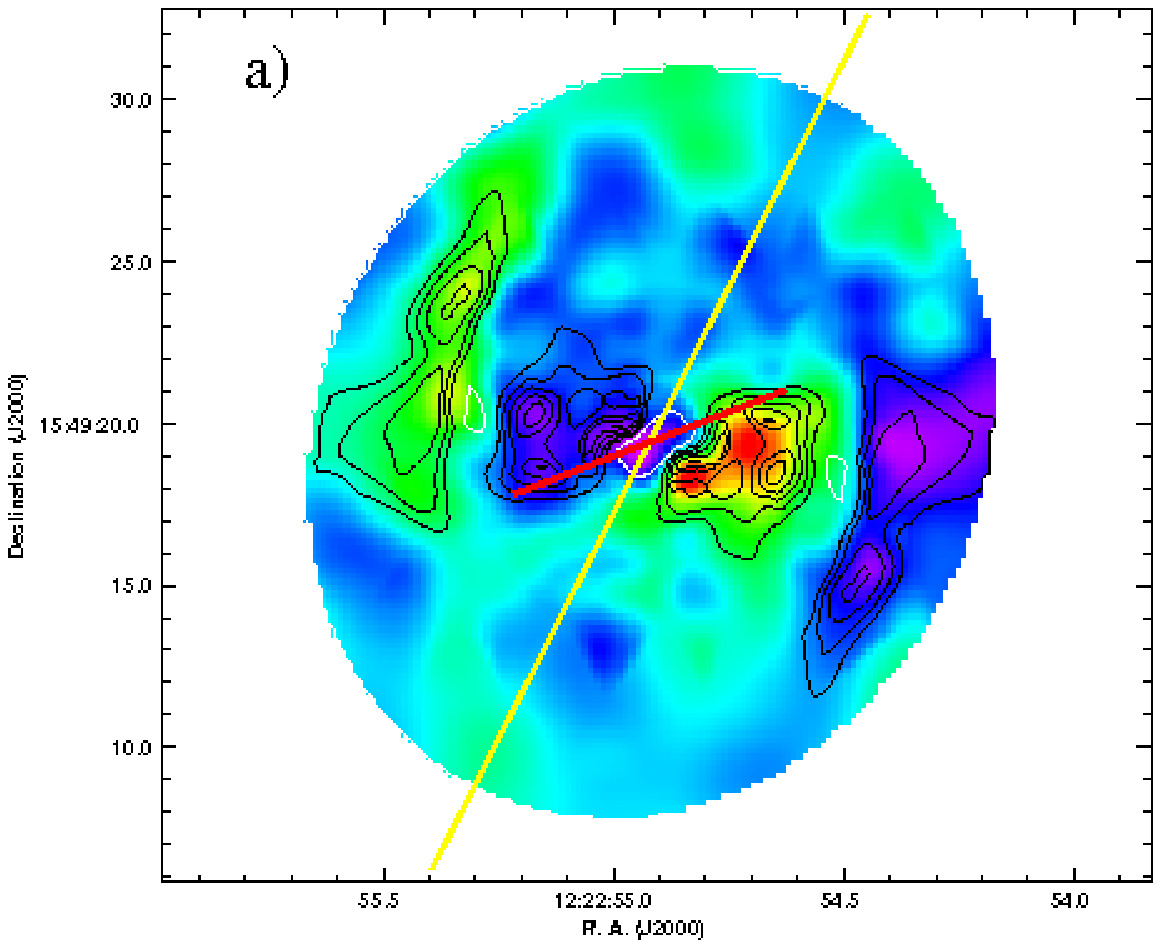}
\includegraphics[angle=0,width=8.4cm, clip=true]{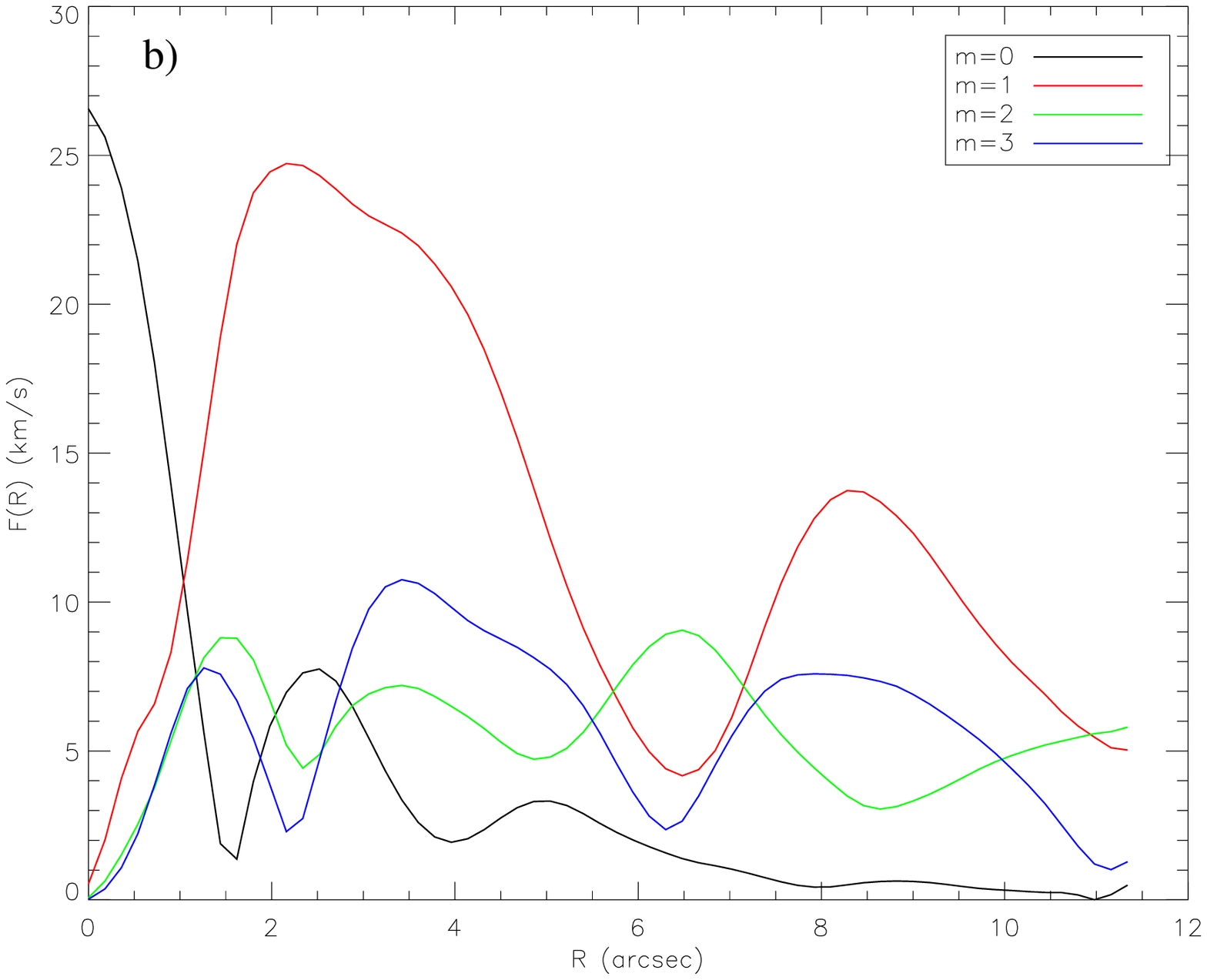}
\caption{{\bf{(a)}} Residual velocity field (colour image) with the
symmetric and
antisymmetric dominance regions overplotted on white and black contours
respectively. {\bf{(b)}} Radial variation in amplitude of the Fourier
modes of
the
residual velocity field. We show the results for the first four modes.}
\label{figure:simasim_fourmodes}
\end{figure*}

It is worth noting that the  highest H$\alpha$ intensity spots take place in
regions of low residual velocities~\citep[see also][]{zurita04}. In fact, in
the case of NGC 4321, the  H$\alpha$ ring of star formation actually lies in
the transition region between   the bar and arm regions (with a radius
of about
6$\arcsec$) with very low  residual velocities. It is indeed very likely
that
both facts (high H$\alpha$ intensity and low residual velocities) are
 closely
related and star formation takes place in a {\em kinematically quiet}
 region.

A careful look at these residuals also reveals that the highest velocity
gradients (which trace shocked gas regions) take place at the inner rim
of the
spiral arms and around the nucleus of the galaxy (at its far side). This is
better illustrated if we calculate the gradient of the residual velocity
field.
We have used the Roberts~\citep{roberts65} operator to compute the 2-D
spatial gradient measurement on the residual velocity field. The residual
velocity gradient contours are shown on top of an
unsharp-masked HST image of NGC 4321 in
Figure~\ref{figure:resco_robdust} (b). It
is quite
impressive how the shocked gas traced by the high gradient matches very
well the
location of the dust lanes along the inner rim of the arms (particularly
for the
northern arm). This indicates that
the spiral arms are very probably formed by hydrodynamic shocks
originated by gas
orbit
crowding at a resonance of the bar potential (as the mass distribution
as traced
by the NIR light
shows no relevant overdensities at these locations). This coincidence
of the dust location with high velocity gradients was
also found and studied in detail for NGC 1530 by~\citet{zurita04}.

The nuclear velocity gradient, although mixed with some bar-generated
shocks,
may also be understood in the context of the presence of an outflow in the
nuclear region. If the approaching side of an outflow (more likely to be
detected than the receding one, located behind the galactic disc)
produces a
region of shocked gas over the nuclear region, this would be seen in
projection
somewhat displaced along the minor axis on the far side of the disc.
Additional
indications of an outflow can be found in the morphology of the emission
line
profiles of the innermost region spectra that show blue shoulders and
blue wings, especially in the high excitation lines ([NII], [SII]). In other galaxies (see e.g. \cite{Arribas_Mediavilla_93}, \cite{Arribas_Mediavilla_94}), these features are an evidence of streaming gas. 
Finally, the existence of a nuclear outflow is also supported by the
blue-shifted interstellar contamination of the Na D absorption (see
Fig.~\ref{figure:sp_gx}(b)), very common in superwind
galaxies~\citep{Heckman00}.

\subsection{Fourier Analysis of the Residual Velocity Field}
\label{section:sym}

We would like to go one step further in the analysis of the residual
velocity
field of NGC 4321.
One of the most striking features of the residual velocity map is its great
degree of symmetry. ~\citet{knapen95a} already noticed that the $K$-band
image of
the nuclear region of NGC 4321 presented a high degree of symmetry. They
performed a decomposition of the symmetric and antisymmetric components
of this
image, finding that an impressive 95\% of the flux is emitted in the
symmetric
component. If, as expected, the $K$-band image is a good tracer of the mass
density and hence of the potential, it is reasonable also to expect  a high
degree of symmetry in the kinematical signature of such a potential. We have
therefore performed the same kind of decomposition in our residual velocity
field. The result of such a procedure is shown in
Fig.~\ref{figure:simasim_fourmodes} (a), where the residual velocity map
is shown
in colours with regions of symmetry and antisymmetry dominance plotted
on white
and black contours respectively.
The three distinctive regions mentioned in Section~\ref{section:gaskin}
now stand
out  more clearly. The central region is the only one that shows a relevant
symmetric dominance (with blue-shifted velocities) while the bar and arm
influence zones are regions of antisymmetric dominance. The rest of the
map does
not show any definite symmetry.

It is not surprising that most of the residuals belong to the antisymmetric
component. As~\citet{knapen95b} showed, there is a clear $m=2$ component
in the
potential generated by the nuclear
bar.~\citet{canzian93}, and later~\citet{schoen97} in an analytical way,
showed that
a perturbation of harmonic number, $m$, in the potential shall generate
$m-1$ and
$m+1$
signatures in the velocity field (dominating inside and outside corotation
respectively).

The antisymmetric component is dominated by an harmonic $m=1$ mode,
while the
symmetric component is contributed by a combination of even modes (mostly
$m=0$).
To show this in a clearer way, we have also performed a Fourier analysis
of the
residual velocity field. The residual velocity field is decomposed into
its Fourier
components in rings of different radii.
Each ring is therefore characterized by the amplitude $F_m(R)$ and phase
$\phi_m(R)$ of each Fourier
component $m$.
 The results are shown for the first four modes in
Figure~\ref{figure:simasim_fourmodes} (b).
The amplitude of modes $m > 4$ is negligible at all radii.
The three regions of definite symmetry coincide with the regions of high
residuals already mentioned: $m=0$ dominance for the nuclear region, and
$m=1$
for the bar (with a very constant phase angle, and spiral arms
(with a phase angle varying with radius). The  $m=3$ mode is second in
importance
in the bar and
spiral arms regions.

\begin{figure}
\centering
\includegraphics[angle=0,width=8cm, clip=true]{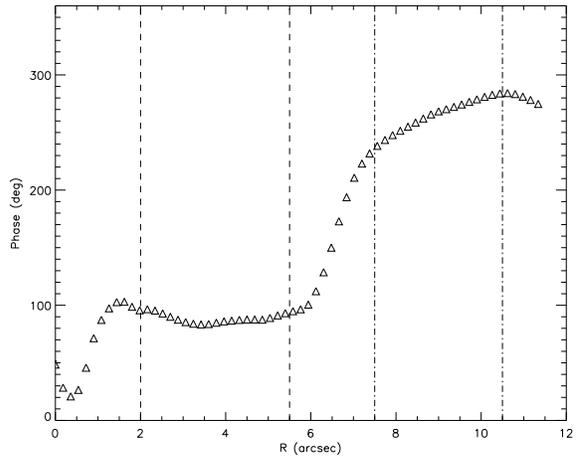}
\caption{{\bf{Phase angle for m=1 mode in the Fourier analysis of the gas residual velocity field. The bar region (between dashed vertical lines) shows a constant phase and the spiral arms region (between dot-dashed vertical lines) shows a phase varying with radius.}} }
\label{figure:phase}
\end{figure}

In our case,
if the bar $m=2$ perturbation to the potential is responsible for the
deviations
from circular rotation, we  expect, according to ~\citet{schoen97} a strong $m=1$ signature (with a secondary $m=3$) in the residual velocity field
(as we are inside the corotation radius and $\Omega_b < \Omega$).
This is exactly what is observed in our data for  NGC 4321.
Although a spiral structure is also a m=2 distortion and would 
therefore produce m-1=1 and m+1=3
signatures in the velocity field, we strongly favour the bar scenario.
The constancy of the phase angle of the m-1=1 in the innermost region
(see Figure~\ref{figure:phase}) clearly indicates a perturbation with a constant PA,
as one would expect from a bar. Moreover, the NIR
photometry of NGC 4321 clearly shows the nuclear bar, but no trace of 
spiral arms in the mass distribution.

The Fourier analysis therefore shows clearly that the velocity residuals are
not local perturbations but {\em global} modes (due to the high degree
of symmetry) generated
by the $m=2$ potential perturbation of the nuclear bar. The spiral arms
are also
generated by the bar potential. The generation of the spiral arms is
most probably caused by shocks originated by gas orbit
crowding at a resonance~\citep[see for example][]{Saka95}.

\subsection{Stellar Rotation Curve. Application of the 
Tremaine--Weinberg Method}
\label{section:stelkin}

We calculate the stellar rotation curve from the
stellar
velocity map (see Fig.~\ref{figure:vel_stellar} (a))  using the same
method employed for the ionized
gas. We fix the position of the centre, inclination, P.A. and systemic
velocity to the values used for the ionized gas, and leave the circular velocity
as the only free parameter in the fit. 
Fig.~\ref{figure:vel_stellar} (a) may give the wrong impression of an offset between the kinematical and photometrical centers. This is due to an combination of facts, the most important being that the systemic velocity plotted in Fig.~\ref{figure:vel_stellar} (a) is the one calculated for the gas
(which is used in the rotation model). The systemic velocity computed from the stellar velocity field 
would give a slightly lower value, resulting in a iso-velocity contour closer to the photometrical center. 
We have chosen to keep the determination of the systemic velocity from the ionized gas (with a much higher S/N) despite this spurious effect.

The resulting stellar rotation curve is
shown in Figure~\ref{figure:vel_stellar} (b). It rises more slowly in the
innermost region but shows a very similar behaviour to the gaseous rotation curve. The
values of the stellar rotational velocity are always smaller than the gaseous ones.
Qualitatively, this is to be expected owing to the stars' asymmetric drift.

\begin{figure}
\centering
\includegraphics[angle=0,width=8cm, clip=true]{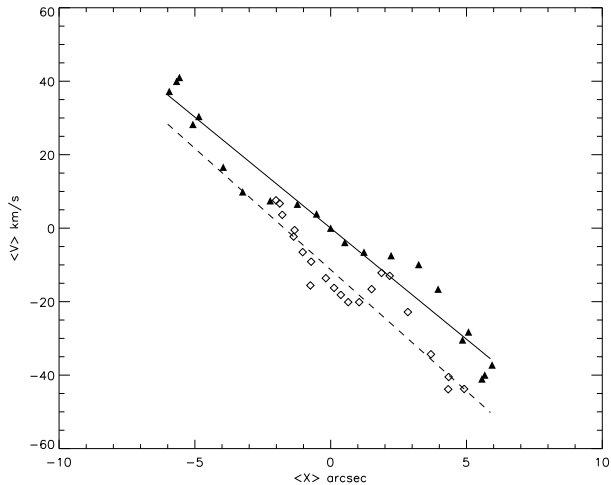}
\caption{ Results of the TW method. This plot shows the results for the
unweighted data (empty diamonds) with the best fit (dashed line)). The {\em
symmetrized} version is shown in solid triangles with the best fit as a
solid line
(see text for details). Both fits (for the unweighted and symmetrized data)
provide very similar results with in a value
of the slope of $\Omega_b$=160$\pm$70 $\rm km \ s^{-1} \ kpc^{-1}$.}
\label{figure:tw}
\end{figure}

\begin{figure}
\centering
\includegraphics[angle=0,width=8cm, clip=true]{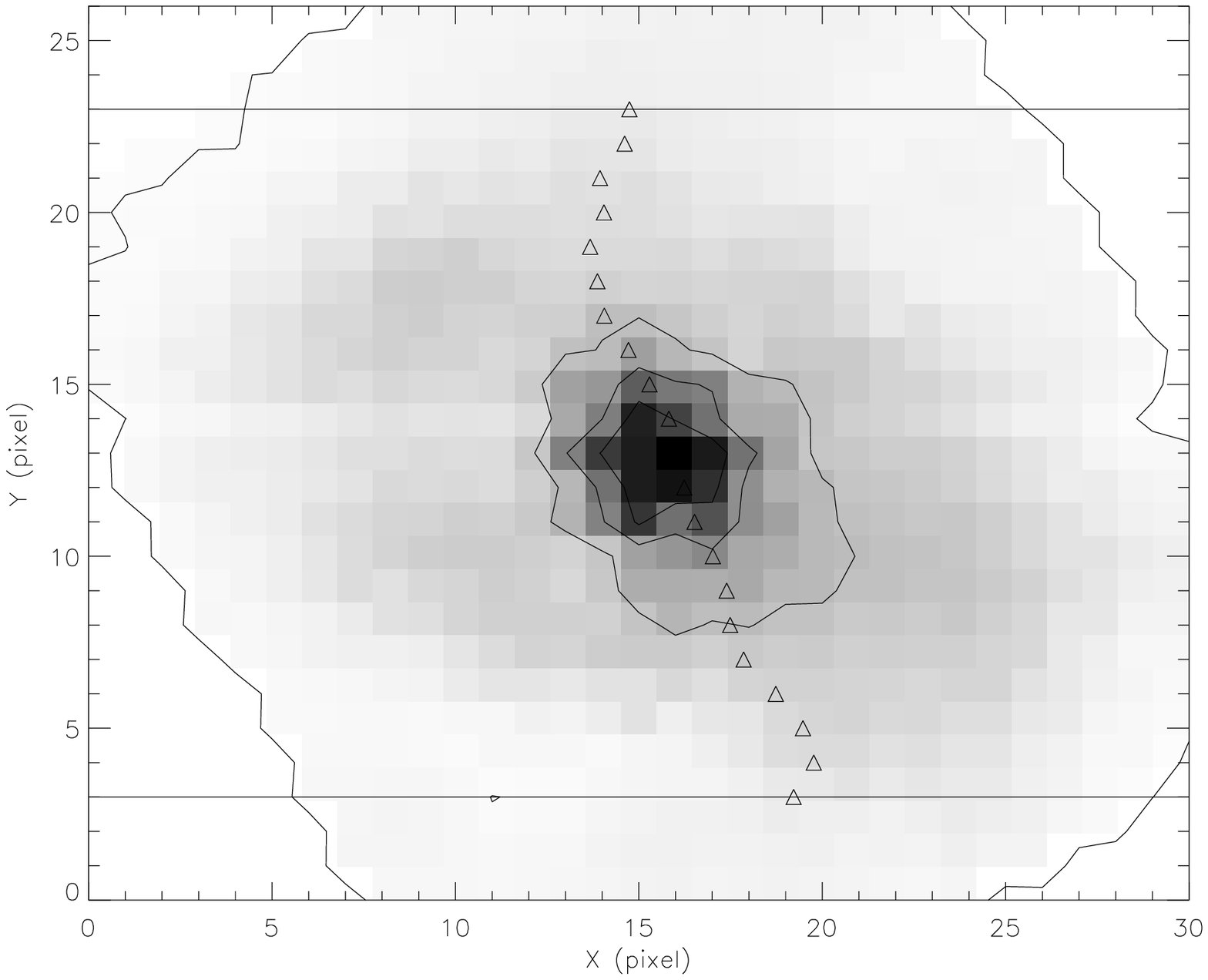}
\caption{{\bf{$<Y>$ vs $>X>$ plot on top of an image of the NaD line strength.}}}
\label{figure:xyplot}
\end{figure}

The pattern speed of the bar in NGC 4321 and the existence of two nested
bars
or a single bar have been a much debated issue~\citep[see][for a
review]{hernandez05}. In any case, what seems to be clear is
that there are several relevant pattern speeds in NGC 4321. Most
observational
determinations~\citep[see][and references therein]{hernandez05} seem to
agree on
a value of
$\Omega_p\approx$ 20-35 $\rm km \ s^{-1} \ kpc^{-1}$ for the outer
spiral arms.
The discrepancies arise in the inner region and over
whether the large scale bar and the nuclear bar are the same or different
structures.
Comparison of numerical simulations with observations
have left two surviving alternatives. On one side~\citet{wada98}
and~\citet{knapen00} favour a single bar scenario in which the pattern
speed of
the bar is $\Omega_b\approx$ 70 $\rm km \ s^{-1} \ kpc^{-1}$. If the
large scale
bar has to rotate at such a high speed, this scenario seems to
contradict the
above-mentioned fact that most observational studies determine a much lower
pattern speed for the outer
regions (where both the large scale bar and spiral arms coexist). The
alternative model proposed by~\citet{Burillo98} consists of two nested
bars with
different (although closely linked) pattern
speeds. Their best solution provides fast and slow pattern speeds of
$\Omega_f$=160 $\rm km \ s^{-1} \ kpc^{-1}$ and $\Omega_s$=23 $\rm km \
s^{-1} \
kpc^{-1}$ respectively. This alternative has the virtue of being able to
drive
gas inwards thanks to the fast mode (as in the single bar model) in the
inner
region while keeping the observed slow mode at large radii.

The best observational determination of the pattern speeds of structures in
galactic discs is provided by
the method proposed by~\citet{tw84} (hereafter TW). This method does not
rely on any particular theory or numerical simulations. It relies
exclusively on
the continuity equation. In a cartesian coordinated system $(x,y)$ where
the $x$- and
$y$-axes are aligned with the galaxy's major and minor axis
respectively, they
showed that
if one can find observationally a tracer of the mass density and the
velocity of
a galactic component that satisfies the continuity equation, then the
underlying
pattern speed can be calculated from this observational information as:
\begin{equation}
\Omega_p=\frac{1}{\sin i} \frac{\int_{-\infty}^\infty \Sigma (x)
(v_{los}(x)-v_{sys}) dx}{\int _{-\infty}^\infty \Sigma (x) (x-x_c) dx} =\frac{1}{\sin i} \frac{\langle V_{los}\rangle}{\langle X \rangle}
\label{eq:tw}
\end{equation}
where $\Sigma (x)$ is the observed intensity (assumed to be proportional
to the surface mass density), $v_{los}(x)$ is the
line of sight velocity, $v_{sys}$ is the systemic velocity and $x_c$ is
the $x$
coordinate of the centre. The integrals are performed along slits or slices
parallel to the major axis direction. 
TW also pointed out
the convenience of using an odd weighting function in $y$ by using, for
example,
slits offset by the same
amount on both sides of the centre.~\citet{merrifield95} refined the
method by
noting that a plot of $\langle V_{los} \rangle$ vs $\langle X \rangle$
provides
a combined measurement of the pattern speed using several slits, which is
insensitive to errors in the centre position and systemic velocity.

As commented above, the TW method has been applied to several gaseous components, such as $\rm H\alpha$~\citep{hernandez05} or molecular gas~\citep{rand04}. However, it is arguable whether these components satisfy the continuity equation. Stars, particularly old stars, on the contrary, are assumed to survive several
cycles around the centre and thus, are expected to satisfy the continuity
equation. It is, therefore, interesting to use our stellar velocity field and the strength of a stellar absorption line (as a tracer of the mass surface density) to apply the TW method to the inner region of NGC 4321. The integrals in Eq.~\ref{eq:tw} are performed along 21 slits parallel to the major axis and centred on the galaxy centre. The slits are offset by 1$\arcsec$. The resulting $\langle V_{los} \rangle$ vs $\langle X \rangle$ plot is shown in Figure~\ref{figure:tw}.
The plot shows also a {\em symmetrized} version by using an odd $\delta(y-y_c)-\delta(y+y_c)$ weighting function in the integrals. 
The $<Y>$ vs $<X>$ plot shown in Figure~\ref{figure:xyplot} removes the possible degeneracies between different pattern speeds~\citep[see][]{hernandez05}.

Both fits (for the unweighted and symmetrized data) provide very
similar results with a slope value of $\Omega_b$ = 160$\pm$70 $\rm km \ s^{-1} \ kpc^{-1}$. The quoted error contains the contribution from uncertainties in the P.A. As a cross-check we have also used  the $K_{\rm s}$ infrared image of
NGC 4321 of~\citet{knapen03} as the mass tracer for the old stars.
The results in this case provide a somewhat higher value for $\Omega_b$
although compatible with the previous one within the large error bars.

This result is very different from the one associated to the single bar scenario ($\Omega_b\approx$ 70 $\rm km \ s^{-1} \ kpc^{-1}$). However, in spite of the numerical coincidence the agreement with the fast mode of the nested bar scenario proposed by~\citet{Burillo98}, is not conclusive.

A linear trend of the $\langle V_{los} \rangle$ vs $\langle X \rangle$ relationship like the one observed in Figure~\ref{figure:tw} is expected for the bar rotation pattern but it is not exclusive of it. In fact, the stellar rotation curve of the inner region of NGC 4321 is approximately described by a linear relationship of about 100$\pm$20 $\rm km \ s^{-1} \ kpc^{-1}$ between 3 and 12 arcsecs. Thus it would be not surprising if the convolution (see Eq.[1]) of the corresponding solid-body like rotating velocity field with a bar of finite width will produce a $\langle V_{los} \rangle$ vs $\langle X \rangle$ linear trend with $\Omega_{rot} \sim 100$ $\rm km \ s^{-1} \ kpc^{-1}$. A direct test repeating the complete procedure to apply the T-W method but now to a synthetic velocity field generated with the stellar rotation curve of Figure~\ref{figure:vel_stellar} yields a value $\Omega_{rot} = 130\pm 30$ $\rm km \ s^{-1} \ kpc^{-1}$. Thus, $\Omega_{rot}$ and $\Omega_b$ = 160$\pm$70 $\rm km \ s^{-1} \ kpc^{-1}$ are compatible within errors. In fact, owing to the relative closeness between these two magnitudes (if the two nested bars scenario is correct) a robust determination of $\Omega_b$ would need a strong diminution of the errors.

This new determination of the pattern speed has obviously an 
impact on the location of the important resonances. A higher pattern speed
moves most resonances to smaller radii. We have used the linear epyciclic
approximation to make an estimation of the resonance location. 
There are two Inner Lindblad Resonances located at radii of approximately 6 and 12 arcsec 
respectively. The Ultra Harmonic Resonance ($\Omega-\kappa/4 = \Omega_b$) is located at a radius 
of about 14 arcsec.
The star forming ring seen in $\rm H\alpha$ is located right inside the 
Outer Inner Lindbald Resonance and Ultra Harmonic Resonance. Whether this star forming ring is related to some of these 
resonances is indeed a possibility, but we cannot be conclusive in
this respect.
In any case, these values must be considered very cautiously, 
as they depend
critically on the shape of the rotation curve in its innermost region,
and on the validity of the linear epyciclic approximation, 
which is probably a poor one in these regions.

A few words of caution about the determination of the bar pattern speed
must be said at this point.
The original TW method eliminates some undesired
integrals of spatial derivatives by using the fact that the integrals
extend to
infinity, where the density
goes to zero. We obviously cannot fulfil this requirement, as  most
observations  having a
limited field of view can not. It is usually assumed that if the
intensity is low enough
at the borders of the
field of view, the method is still usable.
As we have seen before, there seems to be evidence that in NGC 4321
there are several speed patterns that are relevant at different radii. If this is the case,
integrating along a slit that
runs over the two pattern speed regions may contaminate the results of
the TW
method (for the inner, faster mode), even if we select the
appropriate region along the $y$-axis.
In fact, the TW method can be generalized for a continuously varying
pattern
speed with radius~\citep{merrifield06} although its application requires
the
solution of a Volterra integral equation, which is much
more complex and needs very high quality data over the whole region of
interest.
It would then be
necessary to have high S/N data of the
type presented here covering the whole disc of NGC 4321 to perform such
an analysis
and to  
solve this issue definitively.

\section{Summary}
\label{section:summary}

We have performed a Fourier analysis of the residual velocity field of
the ionized gas and determined the stellar
bar pattern speed to study the departures from regular rotation and how
they relate to the gravitational potential.
The main results of this work can be summarized in the following points:
\begin{enumerate}
\item The ionized gas kinematics presents, on top of the galactic
rotation, three clearly distinctive
regions: a) A small nuclear region of blue-shifted velocities.
b) A bar-dominated region where the residuals  originate from the gas
orbits
aligned with the bar major
axis. c) A region dominated by the streaming motion across two symmetric
spiral
arms.
\item The analysis of the gas kinematical symmetries shows these three
regions
in a very clear way as
regions of dominance of symmetry or antisymmetry, or, alternatively, of
$m=0$
and $m=1$ (secondary $m=3$)
Fourier modes. This is exactly what theory predicts if they are
signatures of an
$m=2$ perturbation
of the potential. Thus, the outstanding morphological and kinematic
perturbations of the circumnuclear region of
NGC 4321 present the required (high) degree of symmetry to be explained
by the $m=2$ potential of a bar lying in this region.  
\item We have detected this (otherwise hidden) bar at
optical wavelengths by means of the 2D distribution of the equivalent
width of
the stellar absorption lines corresponding to the old stellar
population. This optical bar matches well the inner bar found in the
near-infrared by Knapen et al.\ (2003).

\item We apply the Tremaine--Weinberg method to the
stellar data to estimate the pattern speed of the
inner bar. The obtained value of $\Omega_b=160\pm70
\ \rm km \ s^{-1} \ kpc^{-1}$ is very different
from the value proposed in the simple bar
scenario.  However, the uncertainties in the
pattern speed determination prevent the
confirmation of ~\citet{Burillo98} results based in
the existence of two nested bars.

\item The nuclear blue-shifted region is probably the
signature of a nuclear outflow. This is supported by the blueward
asymmetry of the
emission line profiles and by the presence of a blue-shifted
interstellar gas
component in the Na D absorption.

\item Finally, while star formation seems to take place in kinematically
quiet regions, the dust location coincides with high velocity gradients.

\end{enumerate}

\section*{Acknowledgments}
This paper has been supported by the ``Plan Andaluz de Investigaci\'on''
(FQM-108) and by the ``Secretar\'{\i}a de Estado de Pol\'{i}tica
Cient\'{i}fica
y Tecnol\'ogica'' (AYA2000-2046-C02-01, AYA2004-08251-C02-02,
ESP2004-06870-CO2-02).
This research has made use of the NASA/IPAC Extragalactic Database (NED)
which
is operated by the Jet Propulsion Laboratory, California Institute of
Technology, under contract with the National Aeronautics and Space
Administration. We have used observations made with the NASA/ESA Hubble
Space
Telescope,
obtained from the data archive at the Space Telescope Science Institute.
STScI is operated by the Association of Universities for Research in
Astronomy, Inc. under NASA contract NAS 5-26555.
J. Jim\'enez-Vicente acknowledges support from the Consejer\'{\i}a de
Educaci\'on y Ciencia de la Junta de Andaluc\'{\i}a.
A.Castillo-Morales acknowledges the support from Universidad Complutense de
Madrid (AYA2003-01676). We thank J. Knapen and A. Zurita for their help
with
the astrometric calibration of our images.
We also acknowledge the support of the RTN Euro3D: "Promoting 3D
spectroscopy in
Europe".
Thanks to the anonymous referee for valuable comments and suggestions that have contributed to improve this paper.

\end{document}